\documentclass[aps,nofootinbib,preprintnumbers,citeautoscript]{revtex4}
\usepackage{amsmath,amssymb}
\usepackage{enumerate}
\usepackage{graphicx}
\usepackage{amsthm}
 \usepackage{mathtools}
 \usepackage{textcomp}
  \usepackage{braket}
  \usepackage{longtable}
    
\begin{document}
\title{Quadratic Sieve Factorization Quantum Algorithm and its Simulation}

	\author{Amandeep Singh Bhatia$^1$ and Ajay Kumar$^2$\\
	\textit{Department of Computer Science, Thapar Institute of Engineering \& Technology, India} \\
	E-mail: $^1$amandeepbhatia.singh@gmail.com}

	\begin{abstract}
	Quantum computing is a winsome field that concerns with the behaviour and nature of energy at the quantum level to improve the efficiency of computations. In recent years, quantum computation is receiving much attention for its capability to solve difficult problems efficiently in contrast to classical computers. Specifically, some well-known public-key cryptosystems depend on the difficulty of factoring large numbers, which takes a very long time. It is expected that the emergence of a quantum computer has the potential to break such cryptosystems by 2020 due to the discovery of powerful quantum algorithms (Shor's factoring, Grover's searching algorithm and many more). In this paper, we have designed a quantum variant of second fastest classical factorization algorithm named “Quadratic Sieve”. We have constructed the simulation framework of quantized quadratic sieve algorithm using high-level programming language Mathematica. Further, the simulation results are performed on a classical computer to get a feel of the quantum system and proved that it is more efficient than its classical variants from computational complexity point of view.
\end{abstract}
\maketitle



\theoremstyle{plain}
\newtheorem{thm}{Theorem}[section]

\theoremstyle{definition}
\newtheorem{defn}{Definition}[section]
\newtheorem{exmp}{Example}[section]

\section{Introduction}
In the early 18th century, integer factorization was identified as fundamental problem and most important research area in computational number theory \cite{nm}.  After the advent of digital computers, it has been applied in various applications related computing, cryptography and information security. Particularly, it is extremely related with the field of cryptography as finding factors of large integers is difficult for computers \cite{yan2002number}.  Over the last three decades, public key cryptosystems (Diffie-Hellman key exchange, the RSA cryptosystem, digital signature algorithm (DSA), and Elliptic curve cryptosystems) has become a crucial component of cyber security \cite{bernstein2009introduction} \cite{chen2016report}. In this regard, security depends on the difficulty of a certain theoretical problems such as integer factorization or the discrete log problem.

Since its beginning, Mathematicians have been trying to find factors of composite numbers in faster and efficient ways \cite{cm}\cite{dn}. Many algorithms have been devised for determining the prime factors of a given integer such as Trial division, Fermat’s method, Elliptic curve method, Pollard’s Rho method and fastest algorithms: Number sieve, Quadratic sieve and many more. These algorithms are differ in complexity and superiority. Each method has become a stepping stones for the next method. Therefore, it is a very challenging task to design an efficient and effective algorithm for hard computational problem in complexity theory.

Quantum computing is a winsome field that deals with theoretical computational systems (i.e., quantum computers) combining visionary ideas of Computer Science, Physics, and Mathematics. Quantum computing relies upon the quantum phenomena of entanglement and superposition to perform operations. Initially, Feynman \cite{fnman} proposed the idea of quantum computing in 1982 after performing a quantum mechanics simulation on a classical computer. In 1994, Shor \cite{shor} designed a quantum algorithm for calculating the factor of a large number {\it n} with space complexity $O(log~n)$ and time complexity $O(n^2~log log n)$ on a quantum computer, and then perform $O(log~n)$ post processing time on a classical computer, which could be applied in cracking various cryptosystems, such as RSA algorithm and elliptic curve cryptography. Through the impetus provided by Shor's algorithm, quantum computational complexity is an exhilarating area that transcends the boundaries of quantum physics and theoretical computer science. 

In 1981, Pomerance \cite{pom} introduced second fastest factorization algorithm named Quadratic sieve. It is an improvement over Kraitchik's and Dixon's factorization method. In early 1990's, Quadratic sieve method was the most effective and efficient general purpose algorithm. It is the best choice method to find factors of numbers under 100 digits after general number sieve method \cite{pomsmooth}. The running time of number field and quadratic sieve depends on the size of a number (\textit{n}) to factorize \cite{tale}. The fastest General number field sieve algorithm and second fastest quadratic sieve algorithm  works in super-polynomial, but sub-exponential time using a classical computer. But, Shor's algorithm shown a great improvement over these algorithms, which can find factors in polynomial time using a quantum computer. The efficiency of quantum algorithms is based on the "Quantum Fourier transform" i.e quantum variant of classical discrete Fourier transform, which can be constructed in polynomial time.

In this paper, we have designed a quantum variant of quadratic sieve method by using quantum parallelism and entanglement. The most significant property $\it entanglement$ separates the classical world from the quantum world. It is one of the most central topics in quantum information theory. Quantum entanglement is purely quantum mechanical correlation between two parts of the quantum system \cite{et}. Further, we have examined its simulation on Mathematica tool and compared its complexity measure with its own classical variant and competitive Shor's factorization method.

After introducing some preliminaries concepts in Section 2, following contributions are claimed. In Section 3, overview of classical quadratic sieve method is given. In Section 4, Quantum quadratic sieve algorithm is designed. Section 5 presents a  simulation results and compared with its predecessors, followed by conclusion in Section 6.

\section{Preliminaries and Definitions}

Before we can discuss the classical quadratic sieve and design its quantum variant, some preliminaries and definitions are given in this section. Linear algebra is an essential mathematical tool for quantum mechanics. Linear operators allow us to represent quantum mechanical operators as matrices and wave functions as vectors on some linear vector space. We assume that the reader is familiar with the notation of quantum mechanics; otherwise, reader can refer to quantum computational, quantum computing \cite{wang} \cite{ni} and number theory \cite{nt}.

\begin{itemize}
	\item {\it Residue class} \cite{rcl}: Let m $\in $ N. It is defined as the set of integers that are congruent to some integer {\it a} modulo {\it m}. For any a $\in $ Z, {[\textit{a}]} denotes the equivalence class to which {\it a} belongs, such that  [\textit{a}]= {\{\it a $\in$ Z: n $\equiv$ a mod m}\}.
	\item {\it Quadratic residue} \cite{qr}: A integer {\it a} is said to be quadratic residue {\it mod m} if for coprime integers {\it m, a} with {\it m $>$0}, the congruence has the solution such that {\it x$^2$ $\equiv$ a mod m}. If it does not have a solution, then  {\it a} is said to be a quadratic non-residue. 
	
	\item {\it Legendre symbol} \cite{qr}: Let {\it p} be an odd prime and {\it a} an integer. Suppose that \textit{gcd(a,p)}=1. Then, the Legendre symbol is defined as 
	
	$$	\left  ( \dfrac{a}{p} \right) =
	\left\{
	\begin{array}{ll}
	0~ if~ {\it a} ~\equiv {\it mod~ p}\\
	$1~ if~ {\it a} ~is~ quadratic ~residue {\it mod~ p}$\\
	$-1~ if~ {\it a} ~is~ quadratic ~non-residue {\it mod~ p}$
	\end{array}
	\right \}$$

	In quadratic sieve, we have small prime factors. Therefore, an integer {\it n $\in $ N} is called as {\it B}-smooth for {\it B $\in $ R$^{+}$ }, if it has no prime factors greater than {\it B}.
	
	\item {\it Exponent vector} \cite{ev}: Let $p_{i}$ denotes the $i^{th}$ prime, such that
	$$m= \prod_{i} p_{i}^{v_{i}} $$
	
	The exponent vectors is $v(m)=(v_1,v_2, ..., v_i)$. Each entry in \textit{v}(\textit{m}) represents the exponent on the $i^{th}$ prime, where the integer 4 is the first prime, 3 the second and so forth.  
	
	\item {\it Partial measurement} \cite{pm}: In formal way, it is defined as a state vector that is projected onto subspace spanned by the single qubits or quantum registers with probability equal to the square of the projection. In other words, a state vector projected on to the orthogonal subspace spanned by quantum registers with remaining probability. It does not covers the whole system, it just look at the part of the system. Consider a two quantum registers of size {\it n} and {\it m} qubits. 
	$$\ket{\phi}= 
	\sum_{i=0}^{2^n-1}{\sum_{j=0}^{2^m-1} c_{i, j} \ket{i,j}}$$
		
		The probability ${\it pr_j}$ to measure {\it j} in the second register is  
		${pr_j}= \sum_{i=0}^{2^n-1} c_{i, j}^{*} c_{i, j}$.
		And, the new state after the measurement is given as 
		$$\ket{\phi^{'}}= \dfrac{1}{\sqrt{pr_j}}\sum_{i=0}^{2^n-1} c_{i, j} \ket{i,j}$$
		Thus, the new state is given as (normalized) projection onto the respective subspace.
		
	\end{itemize}
	\section{Quadratic Sieve Algorithm}
	
	Quadratic sieve is most extensively used and fastest algorithm to find factors of number less than 130 decimal digits. Since the discovery of quadratic sieve algorithm, it was used to factored 100 to 120 digits long number. Recently, in 1994, it has factored 129 decimal-digit RSA challenge number \cite{fc}. It is based on the classical congruences of squares method i.e. finding the squares whose difference is 0 modulo {\it n}, where {\it n $\in$ N} to be factored. 
	Suppose, a integer {\it n} is an odd composite integer for which we need to find factors. First step in this algorithm is to form factor base. It is a set of primes {\it {S}} such that each element of {\it S} is less than smoothness bound {\it B}. Generally, selection of {\it B} is very crucial part of this algorithm to form factor base. Therefore, it helps in avoiding the unnecessary computations throughout the sieve process.

	Sieve of Eratosthenes is factoring algorithm to generate a list of prime numbers less than smoothness bound {\it B}. The detailed algorithm of Sieve of Eratosthenes can be easily found in \cite{s1,s2}. Its efficiency is $O(\sqrt{n})$. It is found to be useful if someone needs lowest 10,000 primes for some computation. It is used to generate prime numbers in bulk. In case, if we have selected very small value of {\it B}, then there are very limited smooth number to find factors. If it is selected very large, then we need to get more numbers for having linear set. Therefore, the probability that a given integer {\it n} is {\it B}-smooth is calculated as $p_r^{-p_r}$, where $~p_r$=
	$\dfrac{ln(n)}{ln(B)}$. Though, its proof is given in \cite{tp}. Now, we have all the primes labeled as {\it $p_2,~p_3, ...$}. Next step is to calculate Legendre symbol for each prime such that quadratic residue {\it $\left(\dfrac{n}{p_i} \right )=1$}, where {\it i~$\in$~2,~3,~...},  for each prime {\it $p_i$}. 
	
	Second step is to sieve through the polynomial values generated through the sequence  $(x^2-n)$, where $x=\lceil\sqrt{n}\rceil,~\lceil\sqrt{n}\rceil+1,...$ to have {\it B}-smooth values. We can define the sieve interval using optimum value of {\it M} such that [$\sqrt{n}+M,~{\sqrt{n}}-M$], where $M= e^{(ln~n~ln(ln~ (n)))^{3\sqrt{2}/4 }} $. Now, we sieve through the each prime number in factor base such that {\it $(x^2-n)\equiv~0~mod~p$}. It means prime {\it p} surely divides polynomial. Note that, it also sieve through the powers of prime in factor base unless it does not greater than {\it B}. It transform the division process into multiplication. At the end, it forms a list of numbers which indeed a factor by using primes. Now, we have prime factors for each sequence and it can be written as m= \begin{equation*}
	\prod_{i=1}^k p_{i}^{e_{i}}
	\end{equation*}.

	Third step is to transform the prime factors for \textit{k}+1~\textit{B}-smooth values in to corresponding exponent vector, such that for 
$$ \vec{v}(x^2-n)~=~(e_1,~e_2,~...,~e_k)$$
	Further, we reduce the above vector {\it $\vec{v}$} to modulo 2. Here, comes the role of linear algebra to generate row vectors whose sum is equal to zero vector by using Gaussian Elimination.
	$${\it \vec{v}(x_1)+\vec{v}(x_2)+...+\vec{v}(x_k)=\vec{0}}$$
	
	Finally, we are left with ${\it x=x_1~.~x_2~...~x_k~(mod~n)}$ and other variable {\it $y^2$} (i.e. product of $x_i^2-n$) is calculated as
	$${\it y=\sqrt{(x_1^2-n)~.~(x_2^2-n)~...~(x_k^2-n)}~(mod ~n)}$$
	Now, we have desired identity such that ${\it x^2~\equiv~y^2~(mod~ n)}$. And, calculate the greatest common divisors (gcd) of integer {\it n} to have factors such that $f_1=gcd(x-y,n)$ and $f_2=gcd(x+y,n)$.
	Although, detailed explanation of quadratic sieve algorithm and its working with an example can be found in \cite{ev,e1,e2}.

	\section{Quantum Quadratic Sieve Algorithm}
	
	In this section, we provides a quantum variant of above classical quadratic sieve algorithm. Give an integer {\it n}, the algorithm will find the factors of {\it n}  such that {\it $x^2\equiv y^2mod~n$}. Assume that system has two quantum registers Register-1 and Register-2. The algorithm consists of three steps (steps 1 through 3) with step 1.1 and 3 requiring the use of classical computer and the remaining all other steps are executed on quantum computer. In step 1.1, we have used classical sieve of Eratosthenes algorithm to form prime state, which runs in an exponential time. Although, there exists an polynomial algorithm to form prime state by using Grover's algorithm, whose oracle is a quantum variant of Miller-Rabin primality test, which can be use in future \cite{pol}. Here, we begin with briefly describing all the steps of an algorithm.

	\begin{longtable}{p{.1\textwidth}  p{.6\textwidth}}
		
		\hline
		Step 1  & [Initialize] The Register-1 and Register-2 are initialized to zero. Therefore, the state of the registers becomes: 
		$$\ket{\psi_0}=\ket{0}_1\ket{0}_2$$ \\
		\hline
		Step 1.1  & [Determine the primes] The prime numbers in the range {(2,\textit{B})} by using prime counting function on Register-1 are listed. Thus, the total state of the system becomes:
		$$\ket{\psi_p}=\dfrac{1}{\sqrt{\pi(B)}}\displaystyle\sum_{p\in~prime\leq B} \ket{p}_1\ket{0}_2$$
		Thus, we load the first register with an equally weighted superposition of all primes in range less than equal to {\it B}. And, Register-2 with zeros.\\
		\hline
		Step 1.2  & [Prepare factor base] Compute the Legendre symbol on all primes below {\it B}. Thus, apply the unitary transformation {\it $U_p: n^\frac{p-1}{2}~mod~p$} to each prime in the Register-1 and store the result in Register-2. The state of the system is changes as:
		$$\ket{\psi_1}=\dfrac{1}{\sqrt{\pi(B)}}\displaystyle\sum_{p\in~prime\leq B} \ket{p}_1\ket{n^\frac{p-1}{2}~mod~p}_2$$ 
		Now, the states of both registers are entangled with each other. \\
		\hline
		Step 1.3  & [Perform measurement] Partial measurement is performed on  Register-2 for which the calculated Legendre symbol {\it $\left(\dfrac{n}{p} \right )=1$}. It will result in a probability distribution over outcomes of the Register-2: 
		$$P_2,_1=Pr[outcome=1]=\displaystyle\sum_{p\in~prime} |\alpha_p,_l|^2$$
		where, $\alpha_p,_l=1/\sqrt{\pi(B)}$ is the amplitude of prime associated with Legendre symbol 1 in Register-2.
		
		The new state of the system becomes: 
		$$\ket{\psi_2}=\dfrac{1}{\sqrt{P_2,_1}}\displaystyle\sum_{p\in~prime}\ket{p}_1\ket{1}_2$$ 
		In order to form the factor base, we measure the second register for each 1. The state of second register will collapse, but the first register stays in superposition.\\
		\hline
		Step 2  & [Initialize] Registers subscript 3, 4 and 5 are initialized to zeros in order to sieve the sequence {\it ($x^2-n$)} for the interval {\it [$\sqrt{n}+M,~{\sqrt{n}}-M$]} by using optimum value of {\it M}. For simplification, it can be written as [\textit{a}, \textit{b}]. $$\ket{\psi_s}=\ket{0}_3\ket{0}_4\ket{0}_5$$
		[Prepare information for quantum registers] Then, apply the Quantum Fourier transform (QFT) to Register-3 and state becomes:
		$$\ket{\psi_{s_1}}=\dfrac{1}{\sqrt{b-a}}\displaystyle\sum_{x=a}^b \ket{x}_3\ket{0}_4\ket{0}_5$$. \\
		\hline
		Step 2.1  & [Generate sequence] Next step is to apply the unitary transformation on the generated sequence in above step of Register-3 and store its result in Register-4. The state of the system becomes:
		$$\ket{\psi_3}=\dfrac{1}{\sqrt{b-a}}\displaystyle\sum_{x=a}^b \ket{x}_3\ket{x^2-n}_4\ket{0}_5$$ 
		The state $\ket{\psi_3}$ shows more than superpositions of three registers. Now, the states of these registers are entangled with each other. \\
		\hline
		Step 2.3  & [Tensor product of sequence and factor base] Take the tensor product of states $\ket{\psi_2}$ and $\ket{\psi_3}$. Thus, the new state of the system becomes: 
		$$\ket{\psi_{23}}=\ket{\psi_2}\bigotimes\ket{\psi_3}$$
		i.e. 
		$$\ket{\psi_{23}}=\dfrac{1}{\sqrt{P_2,_1}}\dfrac{1}{\sqrt{b-a}} \displaystyle\sum_{x=a}^b\ket{x}_3 \sum_{p\in prime} \ket{x^2-n}_4\ket{0}_5\ket{p}_1 \ket{1}_2 $$ 
		
		Apply, QFT on the Register-4 and state becomes:
		$$\ket{\psi_{24}}=\dfrac{1}{\sqrt{P_2,_1}}\dfrac{1}{(b-a)} \displaystyle\sum_{y=a}^b~ \sum_{p\in prime} \ket{1}_2 \sum_{x=a}^b \ket{x}_3\ket{y^2-n}_4\ket{0}_5\ket{p}_1  $$ \\ 
		\hline
		Step 2.4 & [Factorize the sequence] To compute factors, divide the sequence stored in Register-4 using each prime of factor base (Register-1) iteratively such that
		$$\ket{\psi_{25}}=\dfrac{1}{\sqrt{P_2,_1}}\dfrac{1}{(b-a)} \displaystyle\sum_{y=a}^b~ \sum_{p\in prime} \ket{1}_2 \sum_{x=a,~y\in \it integer}^b \ket{x}_3\ket{y^2-n/p}_4\ket{0}_5\ket{p}_1  $$ 
		Here, we use the Register-4 as an input and output as well. Until, its output is an integer, we keep on increment the Register-5. Hence, there is no loss of information in Register- 4. If needed, we can retrieve it using the Register-3. \\
		\hline
		Step 2.5  & [Observe the state of the quantum computer] Perform partial measurement on Register-4 of above state that equals 1 corresponds to smooth number. The probability of observing the state $\ket{1,x,1,e,p}_{2,3,4,5,1}$ is
		$$P_4,_1=Pr[outcome=1]=\displaystyle\sum_{x=a}^b |\beta_{x, 1}|^2$$
	where, $\beta_{x, l}=1/(b-a)$ is the amplitude of each sequence  associated with  1 in Register-4. Note, the comma here denotes that the registers are entangled with each other and {\it e} denotes the exponent associated with each prime for sequence. Thus, the resultant state becomes:
$$\ket{\phi}=\dfrac{1}{\sqrt{P_2,_1}}\dfrac{1}{\sqrt{P_4,_1}} \displaystyle\sum_{x=a}^{b}
~ \sum_{p\in prime} \ket{1}_2 \ket{x}_3\ket{1}_4\ket{e~mod~2}_5\ket{p}_1  $$ 
And also, perform the modulo 2 on exponents stored in Register-5. \\
\hline
Step 3  & [Matrix processing of state vectors] Represent the quantum Register-5 in form of matrix {\it M} corresponds to its prime value for sequence in above state. Then, perform modulo 2 on matrix. 
Now, we have to look for the vector $(\overrightarrow{v})$ by using Gaussian elimination such that ${\it M.\overrightarrow{v}=\overrightarrow{0}}$.
Further, we take the corresponding values of {\it x} from Register-3 using vector and calculate the other variable {\it y~mod~n} (as given in Section 3).
At the end, compute the greatest common divisors (gcd) to get factors of an integer {\it n}:
${\it f_1}=gcd{\it (x-y,~n)}$ and ${\it f_2}=gcd{\it (x+y,~n)}$. \\
\hline

\end{longtable}

\section{Simulation and its Results}
In this section, we demonstrate a simulation of quantum quadratic sieve algorithm on a classical computer using computational language Mathematica. Its helps us to differentiate the result between quantum computer and classical computer. We have used Mathematica packages "Quantum Computing" (binary qubits, tensor products and tensor powers) and "Quantum Notation" (Kets, bras and other quantum objects in Dirac notation).Therefore, high-level programming language Mathematica consists quantum operators and quantum states which gives us the concept for simulation of quantum algorithm on a quantum computer. Simulation of some quantum algorithms are performed in \cite{qd,qd1}.
	\begin{figure}[!h]
	\centering
	\includegraphics[scale=0.6]{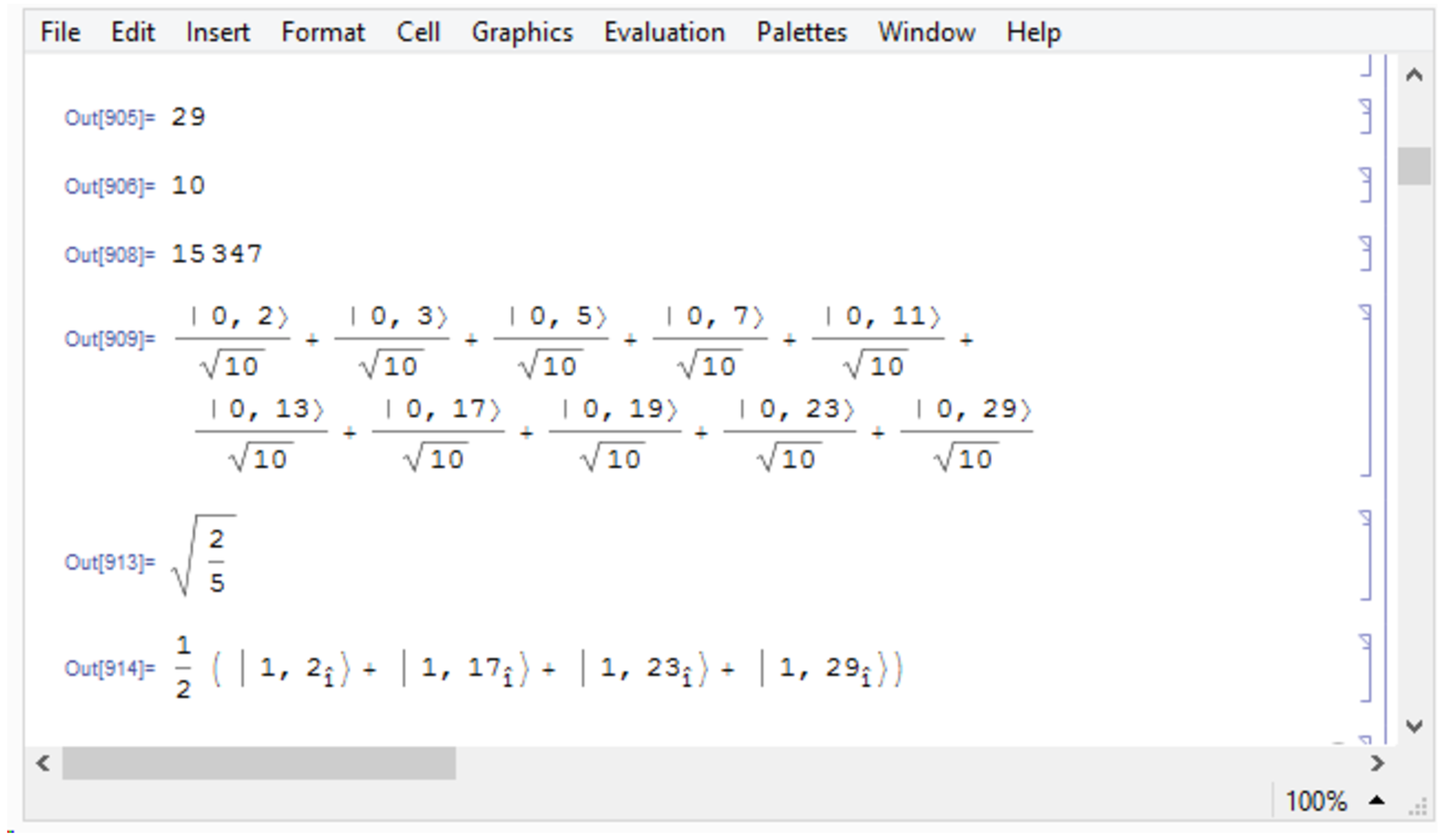}
	\caption{ State after performing partial measurement on calculating Legendre symbol }
\end{figure}

The fastest classical number field sieve factoring algorithm takes   $O(exp(cn^{1/3}(logn)^{2/3}))$ operations for some constant {\it c}, i.e. it is exponential in ${n^{1/3}}$, where {\it n} is the number to factored \cite{lenstra1990number}. Other classical  quadratic sieve algorithm also takes exponential operations to compute factors i.e. $O(exp((1+\epsilon)\sqrt{log n ~log log n}))$, where $\epsilon>$0 \cite{bimpikis2005modern}. Shor's quantum algorithm takes asymptotically $O(n^2logn~loglogn)$, only a polynomial number of operations on quantum computer along with polynomial time on classical computer \cite{ekert1996quantum}. 
On comparing these algorithms complexity with our proposed quantum quadratic sieve algorithm takes $O(n^2logy~loglogn)$ polynomial time on quantum computer, where {\it y} is the number of primes in factor base, classically takes $O(nloglogn)$ to form prime factor base using classical sieve of Eratosthenes and $O(y(w+ylog(y)~log(logy)))$ for Gaussian elimination, where {\it w} is the number of operations needed to multiply with the vector to form null space. In short, if we use trial division method to find smooth numbers in sequence then it will be time consuming process. Thus, by using quantum parallelism, quantum quadratic sieve algorithm shortens the time drastically. Note, its computational complexity can be more improved if we use quantized sieve of Eratosthenes or existing polynomial quantum Grover's algorithm with Miller-Rabin primality test and recently introduced quantum Gauss Jordan elimination algorithm which has computation time of order $O(2^{N/2})$, where $N\geq1$ for ${\it N\times N}$ size matrices \cite{gauss}. Thus, it would be complete quantum quadratic sieve algorithm for factorization taking polynomial time on quantum computer.
	\begin{figure}[!h]
	\centering
	\includegraphics[scale=0.55]{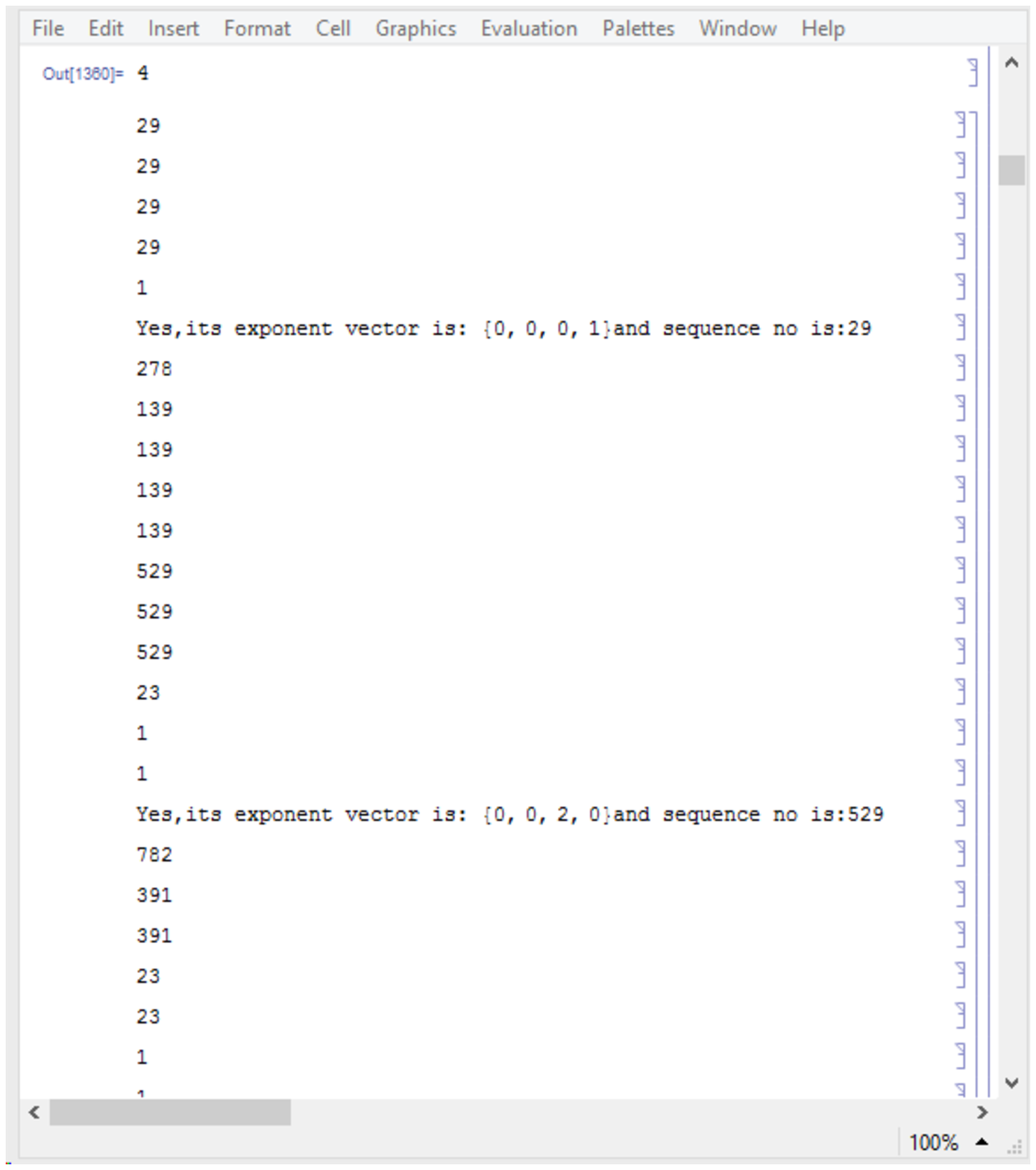}
	\caption{ Generating exponent vectors }
\end{figure}

				However, the proposed quadratic sieve algorithm uses small primes factor base to compute factors efficiently over the sequence, which is computed polynomially over sieving interval. Although, its succession rate depends upon the smoothness bound {\it B}, if it is chosen randomly close to its optimal value, then it would have enough values to form congruent squares and possible to find factors of a large integer in less computational time as compared to its classical variant and number field sieve algorithm. On the other hand, quantum Shor's algorithm is not a factoring algorithm, but rather an algorithm for finding the order of element {\it x} modulo {\it n}, which results in a successful factorization of integer {\it n} \cite{yan2002number}. Therefore, it fails half of the time values of small positive integer \textit{r}, such that $x^r\equiv$1$(mod~n)$.

				In simulation, we have used three gates namely: Hadamard, controlled phase-shift and swap gate to compute QFT. Hadamard gate acts on a single qubit. $\ket{0}=\dfrac{1}{\sqrt{2}}(\ket{0}+\ket{1}) , \ket{1}=\dfrac{1}{\sqrt{2}}(\ket{0}-\ket{1})$.	
				Controlled phase-shift gate modifies the phase of the qubit and its probability of measuring the qubit remains unchanged. Therefore, it maps the $\ket{1}$ to $e^{i\phi} \ket{1}$, where $\phi$ is phase shift and basis state $\ket{0}$ remains as it is. Swap gate exchanges the two qubits i.e. $\ket{10}\rightarrow\ket{01}$.

				$$\ket{\psi_p}=Expand[\dfrac{1}{\sqrt{Z}}\sum_{c=1}^S \ket{Sequence[Part[S,c]],0]}$$

				In step 1.1, the state ${\psi_p}$ is formed by using a function for classical sieve of Eratosthenes which returns an array (\textit{S}) of primes on giving {\it B} as an input. The Expand command expands out a list of primes in Register-1 and zeros in Register-2.		
				Figure 1 lists the all primes with equal probability less than equal to smoothness bound-{\it B}. Also, it shows the probability of measuring a Register-2 in Legendre symbol equal to 1 and outputs the a new quantum state after performing the partial measurement. Note, the comma in between the registers shows entanglement. Figure 2 shows a screenshot of an exponent vectors generated for each sequence dividing with prime factor base.

				\begin{figure}[h]
					\centering
					\includegraphics[scale=0.6]{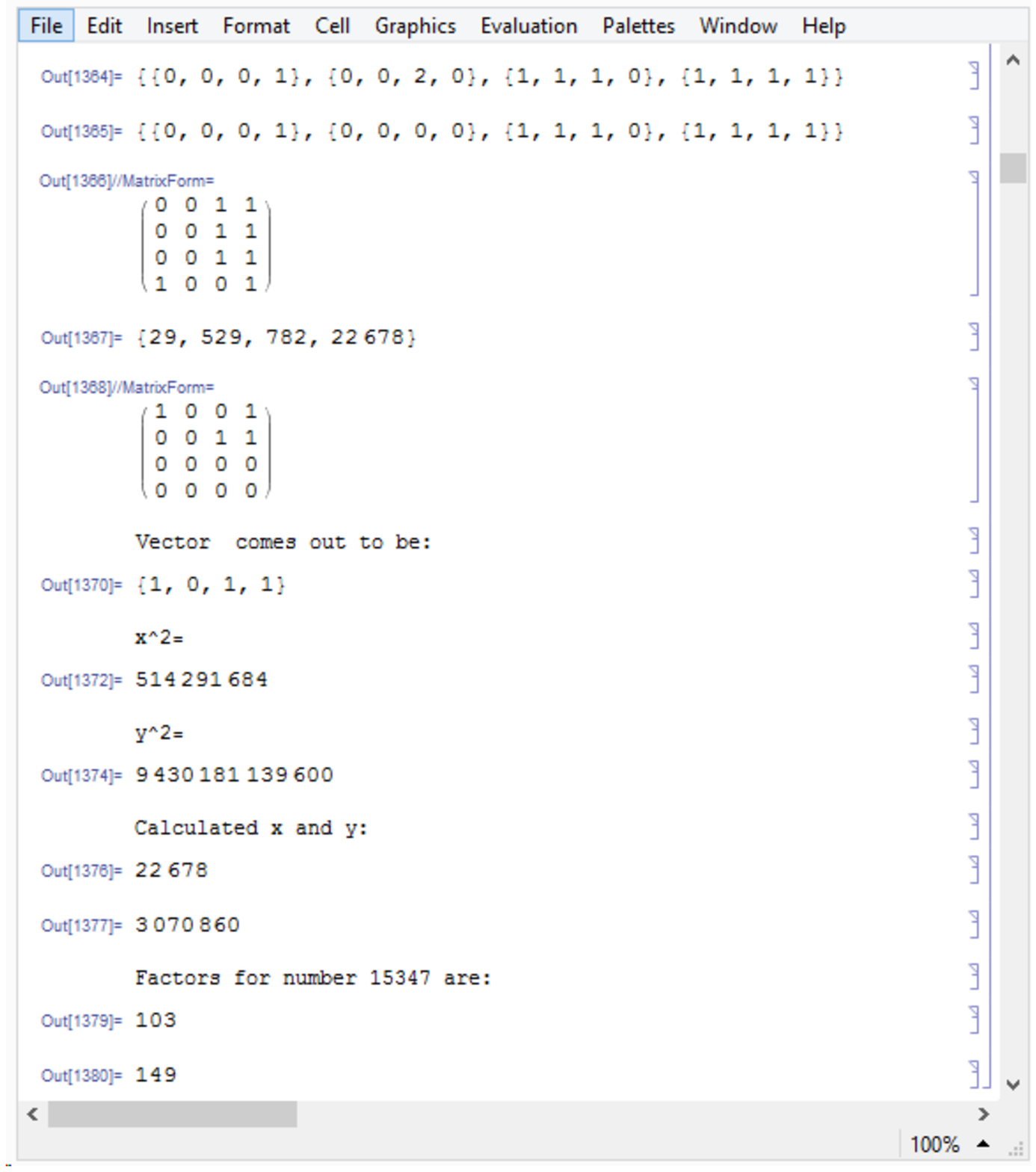}
					\caption{ Simulation results for integer n=15347 }
				\end{figure}
				
				In the last step of algorithm, classical Gaussian elimination is performed on matrix consisting exponent vectors. We have used the QubitToDec command over the quantum states for having decimal values in order to perform arithmetic operations and to calculate greatest common divisors. Finally, we have the factors of an integer {\it n}=15347 i.e. 103 and 149 as shown in Figure 3.

				\section{Conclusion}
				In this paper, we have constructed a quantum version of quadratic sieve algorithm and efficiently simulated it using high-level computational language. It has been proved that proposed algorithm is more efficient in terms of complexity than classical quadratic sieve algorithm. Its efficiency depends upon the quantum Fourier transform which we have used to construct sieving process polynomially. Moreover, we have compared it with the quantum Shor's algorithm and proposed that it can be check in polynomial time if we will use the quantum Gaussian elimination in last step instead of classical Gaussian elimination for large matrices, which is left for future work.  
				
				\section*{Acknowledgments}
				
				Amandeep Singh Bhatia was supported by Maulana Azad National Fellowship (MANF), funded by Ministry of Minority Affairs, Government of India.

\end{document}